# A Transformational Decision Procedure for Non-Clausal Propositional Formulas

Alexander Sakharov
alex@sakharov.net
http://alex.sakharov.net

**Abstract**

A decision procedure for propositional logic is presented. This procedure is based on the Davis-Putnam method. Propositional formulas are initially converted to negational normal form. This procedure determines whether a formula is valid or not by making validity-preserving transformations of fragments of the formula. At every iteration, a variable whose splitting leads to a minimal size of the transformed formula is selected. This procedure performs multiple optimizations. Some of them lead to removing fragments of the formula. Others detect variables for which a single truth value assignment is sufficient. Examples are presented.

## Introduction

Traditionally, decision procedures for propositional logic determine satisfiability of propositional formulas in conjunctive normal form (CNF). The satisfiability problem is dual to the validity problem. Satisfiability of formula A is equivalent to validity of ¬A. It is assumed that non-CNF formulas are turned into CNF first. This way of approaching the satisfiability/validity problem suffers a major drawback as pointed out by many researchers. Normally, propositional formulas are very distinct from CNF. Their structure is such that connectives & ∨ => and ¬ usually stack up one on top of another many times in various combinations. This is a nightmare scenario for conversion to CNF when & sits on top of ∨ or =>, or vice versa. On average, conversion of arbitrary propositional formulas to CNF imposes a significant burden such as introduction of big quantities of new variables and a significant formula size growth.

Humans reason about propositional formulas without converting them into CNF. We always take advantage of the knowledge of formula structure. Conversion to CNF, on opposite, completely destroys the structure, and thus, nulls this advantage. Experiments also confirm that solving propositional formulas in their original form is much faster than conversion to CNF followed by solving the converted formulas in CNF [Sta].

This article introduces a decision procedure that does not involve conversion to CNF. A conversion to so called negational normal form (NNF) is done first instead. As opposed to CNF, the NNF resembles the structure of the original formula. NNF possesses some nice properties characteristic to normal forms. Conversion to NNF does not impose a size growth or introduction of new variables.

Our decision procedure is based on the Davis-Putnam method [DP, DLL] in the sense that it picks a propositional variable and makes two assignments to it: one is true the other is false. It is called splitting. The difference between the Davis-Putnam procedure and other procedures derived from it on one hand and our procedure on the other hand is that ours does not generate two formulas to branch on but transforms the original formula via its specialization. No additional formulas are generated. In this respect, our method is close to deduction procedures.

The decision procedure presented here minimizes the growth of the formula. At every transformation step, it first attempts to apply optimizations that either remove some sub-formulas or confine truth value assignments to one of true/false. After that, it finds a variable whose splitting leads to a minimal formula size growth.

Our decision procedure determines validity of propositional formulas as opposed to the majority of decision procedures for propositional logic. Note that from the perspective of mathematical logic, validity of formulas is of primary interest because it is equivalent to deducibility [Kle]. We will give model-theoretical proofs of the theorems below by reasoning about truth tables and validity [Kle].

## Negational Normal Form

Propositional formulas are built up from variables using connectives ¬ ∨ & and =>. Sometimes connective ~ is also added to the aforementioned primary connectives.

**Definition.** NNF of a propositional formula is defined recursively by the following rules:

1. A literal is in NNF
2. If $A_1,...,A_n$ are either literals or disjunctions, then $A_1$ & ... & $A_n$ is in NNF
3. If $A_1,...,A_n$ are either literals or conjunctions, then $A_1$ ∨ ... ∨ $A_n$ is in NNF

One may look at propositional formulas as trees whose leaves are literals and whose non-terminal nodes represent connectives. Basically, we do not consider connective ¬ applied to a variable as a separate node.

**Definition**. The size of a formula in NNF is the number of nodes in its tree.

**Theorem 1**. Every propositional formula can be converted to an equivalent formula in NNF in linear time. The size of the resultant formula is linear of the size of the source formula. If connective ~ is not used, then the size does not increase after the conversion.

Proof. This conversion is pretty straightforward. It is done in one pre-order traversal by applying the following transformation rules to the formula and then to its sub-formulas:

A&(B&C) -> A&B&C
A∨ (B∨C) -> A∨B∨C
(A&B)&C) -> A&B&C
(A∨B) ∨C) -> A∨B∨C
¬ ($A_1$&...&$A_n$) -> ¬$A_1$∨...∨¬$A_n$ (n ≥ 2)

¬ (A₁∨...∨Aₙ) -> ¬A₁&...&¬Aₙ (n ≥ 2)
¬¬A -> A
A=>B -> ¬A∨B
A~B -> (¬A∨B)&(A∨¬B)

The replacement theorem [Kle] guarantees that the transformed formula is equivalent to the source formula. Since the entire transformation is done within a single traversal in which the time of processing one node is limited by a constant, this transformation can be carried on in linear time. Rules for & and ∨ do not increase formula size. Rules for => and ¬ temporarily introduce additional ¬ nodes that are subsumed later when pushed down to the bottom. Therefore, the resultant size is not more than the original size when connective ~ is not present. Size increase due to unfolding connective ~ is proportional to the number of its occurrences. ♦

The difference between conversion to NNF and CNF is drastic. In contrast to conversion to CNF, conversion to NNF does not introduce new variables, and formula size increase is not an issue.

NNF is monotone in the following sense. If the value of a formula is true under a certain truth value assignment, and if the value of some sub-formula if false under this assignment, then formula's value remains true if to change the value of the sub-formula to true. For convenience in our reasoning, we use constants true and false to denote any valid formula and any unsatisfiable formula, respectively.

## Formula Specialization

The decision procedure presented here specializes formulas in NNF by assigning true and false to one of their variables. Formula T(A,V) defined below is specialization of A for assignment V=true, and F(A,V) is specialization of A for assignment V=false.

**Definition**. For formula A and variable V occurring in A, formula T(A,V) is defined by the following rules:

1. If V's positive literal is a child of a disjunction, then remove the disjunction.
2. If V's positive literal is a child of a conjunction, then remove the literal.
3. If V's negative literal is a child of a disjunction, then remove the literal.
4. If V's negative literal is a child of a conjunction, then remove the conjunction.
5. If a node with only one sibling is removed, then lift the sibling, i.e. replace its parent by the sibling if the sibling is a literal, or replace its parent by several nodes, namely the children of the sibling if the sibling is not a literal.

For formula A and variable V occurring in A, formula F(A,V) is defined by the following rules:

1. If V's positive literal is a child of a disjunction, then remove the literal.
2. If V's positive literal is a child of a conjunction, then remove the conjunction.
3. If V's negative literal is a child of a disjunction, then remove the disjunction.
4. If V's negative literal is a child of a conjunction, then remove the literal.
5. If a node with only one sibling is removed, then lift the sibling.

If A is a propositional formula and B is its sub-formula, then $A_{B|C}$ denotes a formula obtained from A by replacing B by C. Note that we always refer to a single occurrence of B in A. If T(A,V) or F(A,V) is empty, then it is synonymous to true if A is a disjunction, and it is synonymous to false if A is a conjunction. Both T(A,V) and F(A,V) can be generated in linear time of the size of A because this generation can be accomplished within a single traversal of A in post-order, and the time of processing one node is limited by a constant.

**Theorem 2.** If B is a sub-formula of formula A in NNF and variable V does not occur in A outside of B, then $A_{B|T(B,V)\&F(B,V)}$ is valid iff A is valid.

Proof. First of all, in case if B is A, this is a well-known theorem [Qui]. Thus, we only have to prove this theorem for the case that B is not A itself.

Note that for any assignment of truth values to variables occurring in B such that V=true in this assignment, the value of B equals the value of T(B,V). For any assignment of truth values to variables occurring in B such that V=false, the value of B equals the value of F(B,V). Both T(B,V) and F(B,V) are obtained from B by replacing V by true and applying the following equivalence-preserving rules:

¬true -> false
¬false -> true
$A_1$ & ... & true & ... & $A_n$ -> $A_1$ & ... & $A_n$
$A_1 \lor ... \lor$ true $\lor ... \lor A_n$ -> true
$A_1 \lor ... \lor$ false $\lor ... \lor A_n$ -> $A_1 \lor ... \lor A_n$
$A_1$ & ... & false & ... & $A_n$ -> false

Lifting a sibling also preserves equivalence due to associativity rules for & and $\lor$.

Suppose A is valid. Consider an arbitrary assignment α of truth values to variables occurring in $A_{B|T(B,V)\&F(B,V)}$. The value of A is true for α expanded with both V=true and V=false. The value of A under α plus V=true is the same as the value of $A_{B|T(B,V)}$ under α, and the value of A under α plus V=false is the same as the value of $A_{B|F(B,V)}$ under α, that is, these values are true. Since the value of T(B,V)&F(B,V) under α equals either T(B,V) or F(B,V) under α, and the part of $A_{B|T(B,V)\&F(B,V)}$ outside of T(B,V)&F(B,V) is exactly the same as in A, the value of $A_{B|T(B,V)\&F(B,V)}$ under α is also true.

Now suppose $A_{B|T(B,V)\&F(B,V)}$ is valid. Consider an arbitrary assignment α of truth values to variables occurring in A. If V=true in this assignment, then the value of A under assignment α is the same as the value of $A_{B|T(B,V)}$ under α with V=true excluded. The value of $A_{B|T(B,V)}$ is true under this assignment since the value of $A_{B|T(B,V)\&F(B,V\}}$ is true under the same assignment. The only case when the values of T(B,V)&F(B,V) and T(B,V) are different is the following: the value of T(B,V)&F(B,V) is false and the value of T(B,V) is true. In this case, the calculations of the truth values of $A_{B|T(B,V)}$ and $A_{B|T(B,V)\&F(B,V)}$ may not coincide above T(B,V) and T(B,V)&F(B,V) respectively but since only truth tables for connectives & and $\lor$ are involved, changing one of arguments for these connectives from false to true cannot flip the resultant value from true to false.

If V=false in this assignment, then the value of A under assignment $\alpha$ is the same as the value of $A_{B|F(B,V)}$ under $\alpha$ with V=false excluded. Again, the value of $A_{B|F(B,V)}$ is true under this assignment since the value of $A_{B|T(B,V)\&F(B,V)}$ is true under the same assignment. ♦

## Literal Sets

Following [AGOV] and [GGMOV], we associate two sets with every formula A in NNF. The first set $\Delta_0$ contains literals that are implicates of A. The second set $\Delta_1$ contains literals that are implicants of A. Formally, these sets are defined recursively by the following rules.

1. If L is a literal, then $\Delta_0$ (L) = { L }, $\Delta_1$ (L) = { L }
2. if A is $A_1 \vee ... \vee A_n$, then $\Delta_0(A) = \Delta_0 (A_1) \cap ... \cap \Delta_0 (A_n)$, $\Delta_1(A) = \Delta_1 (A_1) \cup ... \cup \Delta_1 (A_n)$
2. if A is $A_1 \& ... \& A_n$, then $\Delta_0(A) = \Delta_0 (A_1) \cup ... \cup \Delta_0 (A_n)$, $\Delta_1(A) = \Delta_1 (A_1) \cap ... \cap \Delta_1 (A_n)$

Given a formula, these sets can be computed for this formula within a single traversal in post-order. The time of processing of every node is proportional to the number of variables in this formula in the worst case.

We adopt major optimizations from [GGMOV]. Our decision procedure works directly with $\Delta_0$ and $\Delta_1$ sets. The decision procedure TAS [GGMOV] works with $\Delta$–trees, and TAS-D from [AGOV] works with $\Delta$^ sets. Both $\Delta$–trees and $\Delta$^ sets are built on top of $\Delta_0$ and $\Delta_1$ sets. Here are two theorems from [GGMOV], which constitute a basis for some optimizations in our decision procedure.

**Theorem 1\***. If literal $L \in \Delta_0$ (A), then L can be deduced from A or equivalently A ~ L & A. If V is a variable and A ~ V & A, then A ~ V & T(A,V). If A ~ ¬V & A, then A ~ ¬V & F(A,V). If $\Delta_0$ (A) contains both positive and negative literals of the same variable, then A is unsatisfiable. If literal $L \in \Delta_1$ (A), then A can be deduced from L or equivalently A ~ L $\vee$ A. If V is a variable and A ~ V $\vee$ A, then A ~ V $\vee$ F(A,V). If A ~ ¬V $\vee$ A, then A ~ ¬V $\vee$ T(A,V). If $\Delta_1$ (A) contains both positive and negative literals of the same variable, then A is valid.

**Theorem 2\***. If B is a child of formula A in NNF and $\Delta_0$ (A) $\cap$ $\Delta_1$ (B) or $\Delta_1$ (A) $\cap$ $\Delta_0$ (B) is not empty, then B is valid in case it is a disjunction, and B is unsatisfiable in case it is a conjunction. If B is a child of formula A in NNF and either $\Delta_1$ (B) $\subseteq$ $\Delta_0$ (A) or $\Delta_0$ (B) $\subseteq$ $\Delta_1$ (A), then A is valid in case it is a disjunction, and A is unsatisfiable in case it is a conjunction.

The second theorem is stated and proven in [GGMOV] in application to $\Delta$-trees. Carrying these proofs onto $\Delta_0$ and $\Delta_1$ sets is very straightforward. The proof of its first statement of theorem 2* is based on the two equivalences: C & (C $\vee$ D) ~ C, C $\vee$ (C & D) ~ C. The proof of the second statement employs De Morgan's laws [Kle]. Note that by the deduction theorem [Kle], if $\Delta_0$ (A) = { $L_1,...,L_n$ }, $\Delta_1$ (A) = { $K_1,...,K_m$ }, then for any $1 \leq i \leq n$, $1 \leq j \leq m$: $K_j$ => A => $L_i$.

## Decision Procedure

**Definition**. Consider the minimal tree (B) containing all occurrences of variable V in propositional formula A in NNF. If B is a conjunction and V's literal occurs in $\Delta_0$ (B) or if B is a disjunction and V's literal occurs in $\Delta_1$ (B), then B is called strict. If B is not strict, all B's children contain occurrences of V, and each B's child contains both positive and negative V's literals in case when B is a conjunction, then B is called complete. IF B is a conjunction, B is not strict, all B's children contain occurrences of V, and B has at least one child in which all occurrences of V are of the same polarity, then B is called semi-complete. Otherwise, i.e. B is not strict, and at least one of B's children does not contain V, then B is called incomplete.

**Definition.** The footprint of variable V in propositional formula A in NNF is the size of V's minimal tree B minus the sum of the sizes of B's children not containing V.

**Definition.** The weight of variable V in propositional formula A in NNF is the sum of the sizes of all sub-formulas containing V or its negation as a child.

**Algorithm TR-VLD**

*Input:* propositional formula A
*Output:* 'valid' or 'not valid'
*Note:* If this algorithm removes a node (or equivalently replaces it by an empty node) and a single sibling is left after this, then the sibling is lifted.

1. If the formula is a variable, then stop - the formula is not valid. Find all propositional variables that occur only positively or only negatively, and then do the following while traversing the formula in pre-order.
If such variable V occurs in a conjunction, remove this conjunction from its parent disjunction. If this conjunction is the top conjunction of the formula, then stop - the formula is not valid. If this variable occurs in a disjunction, then remove all its literals from the disjunction.

2. Calculate $\Delta_0$ and $\Delta_1$ for formula tree nodes and do the following for every node B while traversing the formula in pre-order:
If B is a conjunction (disjunction) and there are literals of the same variable but with opposite signs occurring in $\Delta_0$(B) ($\Delta_1$(B)), then remove B from its parent disjunction (conjunction). If there is no parent and the literals in question occur in a disjunction, then stop - the formula is valid. If there is no parent and the literals in question occur in a conjunction, then stop - the formula is not valid. For every child C of B, if $\Delta_1$ (C) $\subseteq \Delta_0$ (B) or $\Delta_0$ (C) $\subseteq \Delta_1$ (B), then remove B. For every child C of B, if $\Delta_0$ (C) $\cap \Delta_1$ (B) or $\Delta_1$ (C) $\cap \Delta_0$ (B) is not empty, then remove C.

3. Do the following for minimal trees of all variables:
If it is a conjunctive strict minimal tree B for variable V, then remove B from its parent. If B is the entire formula, then stop - the formula is not valid. If it is a disjunctive strict minimal tree B for a positive literal of variable V, then replace B with F(B,V) provided that B has no parent or F(B,V) is a disjunction or literal. If it is a disjunctive strict minimal tree B for a negative literal of variable V, then replace B with T(B,V) provided that B has no parent or T(B,V) is a disjunction or literal. If B has a parent and F(B,V) (T(B,V)) is a conjunction, then add all children of F(B,V) (T(B,V)) as children to the parent of B.

4. While traversing the formula in pre-order, do the following for every node B:

If B is a disjunction and $\Delta_1(B)$ is not empty, then for every literal from $\Delta_1(B)$ such that its variable V occurs more than once in B, replace B by $V \vee F(B,V)$ if this literal is positive, or by $\neg V \vee T(B,V)$ if the literal is negative. If B is a conjunction and $\Delta_0(B)$ is not empty, then for every literal from $\Delta_0(B)$ such that its variable V occurs more than once in B, replace B by $V \& T(B,V)$ if this literal is positive, or by $\neg V \& F(B,V)$ if the literal is negative.

5. Calculate variable footprints and weights. Select a variable with the minimal value of the footprint minus weight. Generate both T(B,V) and F(B,V) for this variable V and its minimal tree B. If B is a conjunction, go to step 6, otherwise (B is a disjunction) go to step 7.

*This value - footprint minus weight - is a reasonable estimate of the change in the size of the formula resulting from the transformation in step 6 or 7 (whichever applies). If this value is negative, than the size presumably decreases. A variable whose splitting will result in a formula of minimal size is selected.*

6. Remove conjunction B from A and replace it by:
a) if T(B,V) or F(B,V) is empty, then nothing replaces B
b) T(B,V)&F(B,V) if each of T(B,V) and F(B,V) is either a disjunction or literal
c) $T_1 \& ... \& T_n \& F(B,V)$ if T(B,V) is $T_1 \& ... \& T_n$ and F(B,V) is a disjunction or literal
d) $T(B,V) \& F_1 \& ... \& F_m$ if T(B,V) is a disjunction or literal, and F(B,V) is $F_1 \& ... \& F_m$
e) $T_1 \& ... \& T_n \& F_1 \& ... \& F_m$ if B is a complete minimal tree, and T(B,V) is $T_1 \& ... \& T_n$, and F(B,V) is $F_1 \& ... \& F_m$
f) $T_1 \& ... \& T_n \& F_1 \& ... \& F_m \& B_1 \& ... \& B_k \& \underline{B}_{k+1} \& ... \& \underline{B}_j$ if B is semi-complete or incomplete, $B_1,...,B_k$ are B's children not containing V, $B_{k+1},...,B_j$ are children of B containing occurrences of V of one polarity, T(B,V) is $T_1 \& ... \& T_{n-j} \& B_1 \& ... \& B_k \& ... \& B'_j$, F(B,V) is $F_1 \& ... \& F_{m-k} \& B_1 \& ... \& B_k \& ... \& B"_j$, and each $\underline{B}_i$ (k+1 ≤ i ≤ j) is $T(B_i,V)$ if the occurrences of V in $B_i$ are negative, $\underline{B}_i$ is $F(B_i,V)$ if the occurrences of V in $B_i$ are positive, one of $B'_i, B"_i$ is $T(B_i,V)$, the other is $F(B_i,V)$
Go back to step 1.

*Note that the last two cases are actually the most common ones because B is a conjunction. $\underline{B}_i$ may represent a conjunction.*

7. Remove disjunction B from A and:
a) if T(B,V) or F(B,V) is empty, then replace B by the remaining formula unless the remaining formula is a conjunction and B has a parent. In the latter case, add all immediate children of the remaining formula as children to the parent of B
b) add $T_1,...,T_n, F_1,...,F_m$ as children to the parent of B if T(B,V) is $T_1 \& ... \& T_n$ and F(B,V) is $F_1 \& ... \& F_m$ where n ≥ 2 and m ≥ 2 (if B does not have a parent, then replace B with $T_1 \& ... \& T_n \& F_1 \& ... \& F_m$)
c) add $T_1,...,T_n, F(B,V)$ as children to the parent of B if T(B,V) is $T_1 \& ... \& T_n$ and F(B,V) is a disjunction or literal (if B does not have a parent, then replace B with $T_1 \& ... \& T_n \& F(B,V)$ )
d) add $T(B,V), F_1,...,F_n$ as children to the parent of B if T(B,V) is a disjunction or literal, and F(B,V) is $F_1 \& ... \& F_n$ (if B does not have a parent, then replace B with $T(B,V) \& F_1 \& ... \& F_m$)
e) add two children to the parent of B: $T_1 \vee ... \vee T_n$ and $F_1 \vee ... \vee F_n$ if B is complete, T(B,V) is $T_1 \vee ... \vee T_n$ and F(B,V) is $F_1 \vee ... \vee F_n$ (if B does not have a parent, then replace B with

T(B,V) & F(B,V) )
f) replace B with $B_1 \vee ... \vee B_k \vee ( ( T_1 \vee ... \vee T_{n-k} ) \& ( F_1 \vee ... \vee F_{m-k} ) )$ if B is incomplete, $B_1,...,B_k$ are B's children not containing V, T(B,V) is $T_1 \vee ... \vee T_{n-k} \vee B_1 \vee ... \vee B_k$, and F(B,V) is $F_1 \vee ... \vee F_{m-k} \vee B_1 \vee ... \vee B_k$
Go back to step 1.

*Again, the two cases when both T(B,V) and F(B,V) are disjunctions are the most common ones because B is a disjunction.*

**Lemma 1**. If formula A contains only positive (only negative) occurrences of variable V, then it is valid iff the formula obtained from A by removing the parents of all V's literals occurring in conjunctions and removing all V's literals occurring in disjunctions is valid.

Proof. Note that this removal is equivalent to assignment V=false if V's literals are positive or assignment V=true if V's literals are negative. Consider the case when all literals in question are positive. If the value of A is true for all assignments in which V is false, then changing it to V=true will never change the value of A to false because only truth tables for connectives & and ∨ are involved (see Theorem 1). The same argument applies to the case of all negative literals. ♦

**Lemma 2**. If U, V are two different variables from B and B ~ U ∨ B, then T(B,V) ~ U ∨ T(B,V), F(B,V) ~ U ∨ F(B,V), T(B,V) & F(B,V) ~ U ∨ T(B,V) & F(B,V). If B ~ U & B, then T(B,V) ~ U & T(B,V), F(B,V) ~ U & F(B,V), T(B,V) & F(B,V) ~ U & T(B,V) & F(B,V).

Proof of all these equivalences is very straightforward. It is done by assuming that one side is true under an arbitrary truth value assignment and proving that the other side is true as well.

**Theorem 3.** Transformations of propositional formulas performed in steps 1, 2, 3, 4, 6, and 7 result in formulas whose validity is equivalent to validity of the source formulas. The time complexity of a single iteration of TR-VAL is O(r·s) where r is the number of variables and s is formula size.

Proof. Lemma 1 guarantees that the resultant formula of step 1 is valid iff the source formula is valid. If sub-formula B is a conjunction and two opposite literals of the same variable occur in $\Delta_0$ (B), then B is unsatisfiable by theorem 1* about literal sets. Similarly, if B is a disjunction and opposite literals of the same variable occur in $\Delta_1$ (B), then B is valid. Therefore, transformations done in the beginning of step 2 lead to replacement of B by an equivalent formula. Transformations at the end of step 2 also lead to replacement of B by an equivalent formula due to theorem 2* about literal sets. Due to the replacement theorem [Kle], the transformations of step 2 preserve equivalence of the entire formula. Note that $\Delta_0$ and $\Delta_1$ still can be used after step 2 because their recalculation could only result in new literals added to some of them.

In step 3, if minimal tree B is a conjunction, then it is removed, which is equivalent to its replacement by T(B,V)&F(B,V) because T(B,V)&F(B,V) is false. T(B,V)&F(B,V) is false because B is equivalent to L & B where L is V's literal. Hence at least one of T(B,V), F(B,V) is false. If minimal tree B is a disjunction and V's literal is positive, B is replaced by F(B,V). In this case, F(B,V) is equivalent to T(B,V)&F(B,V) because. T(B,V) is equivalent to V ∨ B, and hence is true. Similarly, T(B,V) is equivalent to T(B,V)&F(B,V) in case of V's negative literal.

Theorem 2 guarantees that the validity of the resultant formula is equivalent to the validity of the source formula for the transformations of step 3.

Lemma 2 allows us to use $\Delta_0$ and $\Delta_1$ in step 4 without recalculating them. If B was replaced at step 3, then all literals from $\Delta_0(B)$ and $\Delta_1(B)$ remain in $\Delta_0(T(B,V)\&F(B,V))$ and $\Delta_1(T(B,V)\&F(B,V))$, respectively. Therefore, $\Delta_0$ and $\Delta_1$ can be reused for any ancestor node of B. Every sub-formula of a minimal tree replacement is either T(C,V) or F(C,V), where C is a sub-formula of B. Again by Lemma 2, $\Delta_0(C)$ and $\Delta_1(C)$ can be reused for T(C,V) and F(C,V).

Theorem 1* about literal sets guarantees that the transformations of step 4 lead to replacement of sub-formulas by equivalent formulas. Therefore, the transformations of step 4 preserve equivalence of the entire formula.

Transformations performed in steps 6 and 7 can be viewed as replacement of B by T(B,V)&F(B,V) followed by replacement of T(B,V)&F(B,V) or its parent by an equivalent formula. By theorem 2 and the replacement theorem [Kle], the resultant formula is valid iff the source formula is valid.

In cases 6a, 6c, 6d, 6e, the formula replacing B is equivalent to T(B,V)&F(B,V) as a direct corollary of the associativity of &. The resultant formula in case 6f is equivalent to T(B,V)&F(B,V) due to theorem 1*, idempotence of & and the fact that one of B'$_i$, B"$_i$ implies the other. In cases 7a, 7b, 7c, 7d, 7e, the formula replacing B is equivalent to T(B,V)&F(B,V) as a corollary of the associativity of ∨. In case 7f, the formula replacing B is equivalent to T(B,V)&F(B,V) due to theorem 1*, distributivity and idempotence of ∨.

Let r be the number of variables and s be formula size. A single execution of any step of TR-VLD requires a fixed number of traversals of the formula. The time of processing every node is either independent of the size of the input, or proportional to r, or proportional to the number of children. The time complexity of traversals in which node processing time is proportional to r is $O(r \cdot s)$. Since no node is a child to more than one parent, the time complexity of the other traversals is $O(s)$. In these other traversals, the time of processing a single node is either independent of the size of the input or proportional to the number of children. Therefore, a single iteration of the decision algorithm is carried on in $O(r \cdot s)$. ♦

Here are specific time complexities. A single execution of step1 requires one traversal of the formula in pre-order. Its time complexity is $O(s)$. Calculation of $\Delta_0$ and $\Delta_1$ is carried on in $O(r \cdot s)$. Checking conditions in step 2 is also carried on in $O(r \cdot s)$. The time complexity of the transformations of step 2 is $O(s)$. Finding and marking minimal trees in step 3 is carried on in $O(r \cdot s)$, and the time complexity of the transformations of step 3 is $O(s)$. The time complexity of the transformations of step 4 is $O(r \cdot s)$. Calculation of the footprints and weights is carried on in $O(r \cdot s)$. The time complexity of the transformations of steps 6 and 7 is $O(s)$.

In many cases, formula transformations do not increase formula size; the following examples confirm this. If the size of a formula never increases during a decision procedure run, then determination of validity is carried on in $O(r^2 \cdot s)$. Moreover, if there is such constant c that size increase at each iteration of our decision procedure is not more than c, then still determination of validity is carried on in $O(r^2 \cdot s)$.

## Examples

**Example 1.** Consider formula:
(p∨q)=>((p∨q∨r)&(r=>((q=>r)&(¬r∨p∨q))))

Conversion to NNF gives the following formula:
(¬p&¬q) ∨ ((p∨q∨r)&( ¬r∨ ((¬q∨r)&( ¬r∨p∨q))))

After optimizing this formula at step 4, TR-VAL selects r first for splitting. Note that r's minimal tree is not the entire formula. The outcome after splitting r is: (¬p&¬q)∨((p∨q)& (p∨q)). The following execution of step 2 yields 'valid'.

**Example 2.** Let us look at formula:
(q∨r∨¬s)&((q∨r)=>s)&((p=>(q=>r))=>((p=>q)=>(p=>r)))

Conversion to NNF gives the following formula:
(q∨r∨¬s)&(( ¬q&¬r) ∨s)&((p&q&¬r) ∨ (p&¬q) ∨¬p∨r)

Here, p's minimal tree is strict. Single assignment p=true is done at step 3. TR-VAL replaces p's minimal tree by: (q&¬r) ∨¬q∨r

The entire formula becomes:
(q∨r∨¬s)&(( ¬q&¬r) ∨s)&((q&¬r) ∨¬q∨r)

After that, sub-formula (q&¬r) ∨¬q∨r is reduced to q∨¬q∨r at step 4. Variable s is selected next for splitting due to the minimal value of the footprint minus weight. The outcome of splitting s is this formula: (q∨r)&¬q&¬r&( q∨¬q∨r). The following execution of step 2 yields 'not valid'.

**Example 3.** Now consider example 7 from [AGOV]:
(((p=>¬s)=>(q&¬r))&(q=>r))=>((p=>(¬s&q))=>(r&¬s))

Conversion of this formula to NNF gives the following:
((¬p∨¬s)&( ¬q∨r)) ∨ (q&¬r) ∨ (p&(s∨¬q)) ∨ (r&¬s)

Presumably this example should highlight the advantages of TAS-D [AGOV]. At the first glance, it does not seem the best material for TR-VAL because no minimal tree is smaller than the entire formula and overall application of TR-VAL optimizations may not be as impressive as it is in the two previous examples. Still, TR-VAL behaves excellent – every iteration decreases formula size.

Based on the footprint and weight values, TR-VAL selects r first for splitting. The outcome of splitting r is:
((¬p∨¬s∨¬s)&((( ¬p∨¬s)& ¬q) ∨q)) ∨ (p&(s∨¬q))

Step 4 of the next iteration reduces this formula:

$((\neg p \vee \neg s) \& (\neg p \vee \neg s \vee q)) \vee (p \& (s \vee \neg q))$

Next, p is split, which results in: $(\neg s \& (\neg s \vee q)) \vee s \vee \neg q$. The following step 2 yields 'valid'.

## Related Work

Our decision procedure TR-VLD is a variant of the Davis-Putnam procedure [DP, DLL] for non-clausal propositional formula. Let us compare it to other decision procedure using the Davis-Putnam approach and applicable to non-clausal propositional formulas. Out of all other variants of the Davis-Putnam procedure known to the author, TR-VLD is the only transformational one that does not branch. Other variants of the Davis-Putnam procedure recur on two generated formulas after splitting a variable. Instead, our procedure transforms fragments of the original formula when splitting a variable. The size of a formula may grow as a result of this transformation but size increase is often far less than the size of an additional generated formula.

The PTAUT decision procedure by Armando and Giunchiglia [AG] implements the Davis-Putnam method for non-clausal formulas. PTAUT performs two optimizations called Polarity and Top-Level Disjunctive Occurrence. These optimizations correspond to Affirmative-Negative Rule and One-Literal Clause Rule of the Davis-Putnam-Loveland decision procedure [DP,Lov]. They are also called Pure Literal rule and Unit Clause rule, respectively [Ott]. TR-VLD performs these optimizations as well. TR-VLD applies the Top-Level Disjunctive Occurrence optimization to a much broader set of cases, that is, to all strict minimal trees as opposed to applying it to the entire formula only.

Two other decision procedures implementing the Davis-Putnam method for non-clausal formulas were suggested by Giunchiglia and Sebastiani [GS]: DP* and DP**. DP* and DP** actually work with CNF but take into account original non-CNF formulas. They split variables from the original non-clausal formulas only but not the variables introduced in the process of conversion to CNF.

The decision procedures from [AGOV], [GGMOV], and [Ott] are perhaps the closest to our decision procedure because they all are based on the Davis-Putnam method and work with formulas in NNF. The Otten's decision procedure [Ott] utilizes the Pure Literal and Unit Clause optimizations. TR-VLD performs these optimizations as well. In comparison with Otten's decision procedure, the scope of the Unit Clause optimization is expanded in our procedure to all minimal trees. The decision procedure TAS-D from [AGOV] employs a generalization of the Pure Literal rule and an extension of collapsibility introduced in [Gel]. The decision procedure TAS from [GGMOV] utilizes a wide range of optimizations. TAS actually works with $\Delta$–trees built from propositional formulas in NNF. TR-VLD covers major optimizations from [AGOV] and [GGMOV]. The remaining optimizations could be embedded in it if deemed advantageous for our method.

The most significant difference between TR-VLD and the three aforementioned decision procedures is that our decision procedure is transformational. It replaces the minimal tree of the split variable by a new formula. For incomplete minimal trees, only some of the children of the minimal tree are replaced. These transformed minimal trees or their parts are similar to fragments of the new formulas generated by TAS-D, TAS and others. In [AGOV], $\Delta_0$ and $\Delta_1$ sets serve to construct Δ^ sets. In [GGMOV], they serve to construct Δ–trees. In our method, sets $\Delta_0$ and $\Delta_1$ are used directly.

Selection of the best variable for splitting is crucial in TR-VLD. It helps avoid formula size growth in many instances. In particular, formula size decreases if TR-VAL ends up selecting a variable requiring a single truth value assignment. To author's knowledge, the variable selection method of TR-VLD is unique. Selection of a variable whose splitting gives a minimal formula size often counterweights and even surpasses optimizations because calculation of T(...) and F(...) subsumes relevant optimizations.